# New tidal paradigm in giant planets supported by rapid orbital expansion of Titan

Valéry Lainey[1,2*], Luis Gomez Casajus[3], Jim Fuller[4], Marco Zannoni[3], Paolo Tortora[3], Nicholas Cooper[5], Carl Murray[5], Dario Modenini[3], Ryan Park[1], Vincent Robert[2,6], Qingfeng Zhang[7]

[1]Jet Propulsion Laboratory, California Institute of Technology,

4800 Oak Grove Drive, Pasadena, CA 91109-8099, United States

[2]IMCCE, Observatoire de Paris, PSL Research University, CNRS,

Sorbonne Universits, UPMC Univ. Paris 06, Univ. Lille, 77

[3]Dipartimento di Ingegneria Industriale, Università di Bologna, 47121 Forlì, Italy

[4]TAPIR, Walter Burke Institute for Theoretical Physics

Mailcode 350-17, Caltech, Pasadena, CA 91125, USA

[5]Queen Mary University of London, Mile End Rd, London E1 4NS, United Kingdom

[6]IPSA, 63 bis boulevard de Brandebourg, 94200 Ivry-sur-Seine, France

[7]Department of Computer Science, Jinan University, Guangzhou 510632, P. R. China

*To whom correspondence should be addressed; E-mail: lainey@imcce.fr.

**Tidal effects in planetary systems are the main driver in the orbital migration of natural satellites. They result from physical processes occurring deep inside celestial bodies, whose effects are rarely observable from surface imaging. For giant planet systems, the tidal migration rate is determined by poorly understood dissipative processes in the planet, and standard theories suggest an orbital expansion rate inversely proportional to the power $11/2$ in distance[1], implying little migration for outer moons such as Saturn's largest moon, Titan. Here, we use two independent measurements obtained with the *Cassini* spacecraft to measure Titans orbital expansion rate. We find Titan migrates away from Saturn at $11.3 \pm 2.0$ cm/year, corresponding to a tidal quality factor of Saturn of $Q \simeq 100$, and a migration timescale of roughly 10 Gyr. This rapid orbital expansion suggests Titan formed significantly closer to Saturn and has migrated outward to its current position. Our results for Titan and five other moons agree with the predictions of a resonance locking tidal theory[2], sustained by excitation of inertial waves inside the planet. The associated tidal expansion is only weakly sensitive to orbital distance, motivating a revision of the evolutionary history of Saturns moon system. The resonance locking mechanism could operate in other systems such as stellar binaries and exoplanet systems, and it may allow for tidal dissipation to occur at larger orbital separations than previously believed.**

Saturn is orbited by 62 moons, and the intricate dynamics of this complex system provide clues about its formation and evolution. Of crucial importance are tidal interactions between the moons and the planet. Each moon raises a tidal bulge in the planet, and because Saturn rotates faster than the moons orbit, frictional processes within the planet cause the tidal bulge to lead in front of each moon. Each moon's tidal bulge pulls the moon forward such that it gains angular momentum and migrates outward, similar to the tidal evolution of the Earth-Moon system. However, in giant planets such as Saturn, the dissipative processes that determine the bulge lag angle and corresponding tidal migration timescale remain poorly understood.

Prior monitoring of the mid-sized inner moons' orbital locations suggests that they are migrating outward faster than allowed if they formed at the same time as Saturn [3,4]. These observations motivated two new formation scenarios. One possibility is that Saturn's rings have viscously spread over time, steadily forming mid-sized moons at their outer edge defined by the Roche limit [5]. Another possibility is that resonances between the orbits of Saturn's mid-sized moons and the pull of the Sun can lead to collisions within the satellite system, after which new moons conglomerate from the debris disk [6]. While each scenario predicts different ages for the satellites, these prior studies have assumed a constant tidal lag angle $\theta$ for each moon's tidal bulge, parameterized by a tidal quality factor $Q \simeq 1/\theta$. The $Q$ governing the tidal interaction with each moon is inversely proportional to the tidal energy dissipation rate within Saturn [7]. Denoting the semi-major axis $a$, the orbital expansion rate $t_{\text{tide}}^{-1}$ of each moon is

$$t_{\text{tide}}^{-1} = \frac{\dot{a}}{a} = \frac{3k_2}{Q}\frac{M_{\text{moon}}}{M}\left(\frac{R}{a}\right)^5 n, \qquad (1)$$

where $M_{\text{moon}}$ is the mass of the moon, $R$ and $M$ are the radius and mass of Saturn, $k_2$ is the Love number of degree two, and $n = 2\pi/P_{\text{orb}}$ is the moon's mean motion. Because of the strong dependence on $a$, most tidal theories predict slower migration for outer moons such as Titan.

To help explain the rapid migration of the mid-sized moons measured by [3,4], a new paradigm for the tidal evolution of moons, known as resonance locking, was proposed by [2]. Tidal dissipation due to inertial waves in Saturn's convective envelope [8] or gravity modes in Saturn's deep interior [9] is enhanced at discrete resonances with planetary oscillations. The resonant frequencies are determined by Saturn's internal structure, which is slowly evolving due to processes such as gravitational contraction, helium rain out [10], and core erosion [11]. Moons can get caught in these resonances as Saturn's structure evolves, causing the moons to migrate outward on a timescale determined by Saturn's internal evolution. While explaining the fast orbital expansion of Rhea, [2] predicted a similar expansion rate (and smaller tidal Q) for Titan, making the monitoring of Titan's orbit a strong case for testing their model. Contrary to most tidal theories where the tidal $Q$ is constant for all moons, resonance locking predicts the tidal $Q$ for outer moons is much smaller.

To measure the migration rates of Saturn's moon, we use two independent methods. In the first approach, a coherent orbit of Titan was determined by reconstructing the trajectory of the *Cassini* spacecraft during 10 close encounters of the moon between February 2006 and August 2016. During each Titan encounter, we are sensitive to the relative position of the *Cassini* spacecraft with respect to both the moon and Saturn, providing indirect information on the orbit of Titan during the timespan of the *Cassini* mission. Our data sets encompass only radio tracking data acquired by the ground antennas of the Deep Space Network, namely Doppler observables at X- and Ka-band (8.4 GHz and 32.5 GHz, respectively), and range data at X-band. Due to limited temporal coverage of radiometric data in the vicinity of the other moons, it was not possible to obtain a reliable estimation of their orbits, which were instead retrieved from the latest satellite ephemerides released by JPL (see Methods).

The radiometric data analysis strategy was based on the classical approach used by the *Cassini* Radio Science Team in the past for gravity science experiments [12–17]. Our solution was obtained using JPL's orbit determination program MONTE[18], using a linearized weighted least squares filter that allowed us to determine corrections to an a-priori dynamical model taking into account all the relevant accelerations that affected the orbit of Titan and the trajectory of the *Cassini* spacecraft. The least squares information filter used a multi-arc approach, in which radiometric data obtained during non-contiguous orbital segments, called arcs, are jointly analyzed to produce a single solution of a set of global parameters, which affect all the arcs. Our global parameters include the initial state vector of Titan, its gravity field up to the 5th degree and order, Saturn's gravity field up to $J_6$, Saturn's tidal parameters $Re(k_2)$ and $Im(k_2)$ at Titan's frequency, and the *Cassini*'s thermal recoil acceleration.

Our second method is based solely on classical astrometry data. Similar to [4], we used more than a century of observations, starting in 1886 through the whole *Cassini* mission. New observations of the main moons from [19,20] were added to supplement those from [4]. Our model solved the equations of motion of the eight main moons of Saturn, with the addition of the four Lagrangian moons of Dione and Tethys, as well as Methone and Pallene. The Lagrangian moons are useful to obtain Saturn's Love number $k_2$, while Methone and Pallene are very sensitive to Mimas' mass and Saturn's gravity field. The perturbation of the four innermost moons of Saturn, Prometheus, Pandora, Janus and Epimetheus is introduced by ephemerides. We checked that the chaos affecting the orbits of these moons, as well as a possible secular variation of Saturn's $J_2$, did not affect our results (Methods).

In addition to the initial state vectors of the moons, we fitted the masses of the moons and their primary, the $J_2$, $J_4$ and $J_6$ of Saturn's gravity field, the orientation and precession of the Saturn's pole, Saturn's $k_2$ (that assumes $k_{20} = k_{21} = k_{22}$), and the tidal ratio $k_2/Q$ at the tidal frequencies of Mimas, Enceladus, Tethys, Dione, Rhea and Titan. Due to the large uncertainty in Enceladus' current tidal dissipation rate [21,22], we performed three independent fits, assuming a broad range of values of 10, 33 and 55 GW.

Our results are shown in Figure 1 and Table 1. From the radio tracking data, we measure the tidal quality factor of Titan's tidal bulge to be $Q = 124 \pm 22$ ($3\sigma$ uncertainties), assuming a fixed $Re(k_2)$ equal to 0.382. From astrometry, we find a slightly smaller value $Q = 61^{+240}_{-31}$, but the two are consistent within $2\sigma$ uncertainties. We tested the reliability of these results by performing many trials with different parameters (Methods), finding no substantial variation in our result. This unexpectedly small value of $Q$ for Saturn's tidal bulge raised by Titan is much smaller than our astrometric measurements of the $Q$s for the tidal bulges of other moons, which range from $Q \sim 300$ for Rhea to $Q \gtrsim 3000$ for Tethys. Each tidal bulge clearly has a different value of $Q$, and all well-constrained values lie below the minimum value $Q = 1.8 \times 10^4$ predicted if the moons formed at the same time as Saturn and $Q$ is constant [23]. Hence, our results show that most of Saturn's moons, including Titan, are migrating outward more rapidly than expected.

The rapid migration of Titan is unexpected for all tidal dissipation mechanisms, except for resonance locking, which predicted the observed migration. Figure 1 shows the predicted tidal $Q$ for a resonance locking model with inertial waves with planetary spin evolution timescale $t_p = 6$ Gyr (supplementary information), where the timescale $t_p$ is a parameter of the model that is expected to be comparable to the age of the solar system. In this model, the migration timescale $t_{\text{tide}} = a/\dot{a} \approx 3t_p/2$ of each moon is roughly the same and is driven by the rate at which inertial wave "resonant" frequencies evolve along with Saturn's spin and structure. However, the predicted $Q$ is different for each moon, with smaller values for outer moons, similar to the measurements. Despite the different values of $Q$, Figure 2 shows that the observed migration timescales for each of Saturn's moons are indeed very similar, with $t_{\text{tide}} \approx 10$ Gyr. Hence, we interpret our observations as strong evidence that a resonance locking process is driving the migration of many of Saturn's moons. The nearly constant migration time $t_{\text{tide}}$ for several of Saturn's moons suggests that resonance locking with inertial waves, rather than gravity modes (which predicts smaller $t_{\text{tide}}$ for outer moons), is the most probable explanation for the moons' migration (supplementary information). This may also help explain why mean motion resonances between moons have survived, as resonance locking with gravity modes typically results in divergent migration that can disrupt mean motion resonances between moons.

Table 1: **Top:** Retrieved tidal parameters of Titan's tidal bulge, and their associated $3\sigma$ uncertainties, using *Cassini* radio tracking data. **Bottom:** Saturn's Love number $k_2$ and inverse quality factor $1/Q$ governing the migration of each moon, based on astrometric data. To account for tidal dissipation inside Enceladus, we performed four independent fits, assuming a heating rate of 3, 10, 33 and 55 GW. Error bars are $3\sigma$ formal uncertainties.

| Parameter | Value | Uncertainty |
|---|---|---|
| $Re(k_2)$ | 0.33 | 0.20 |
| $Im(k_2)$ | $-3.08 \times 10^{-3}$ | $0.55 \times 10^{-3}$ |

| Enceladus heating rate | 3 GW | 10 GW | 33 GW | 55 GW |
|---|---|---|---|---|
| Saturn's $k_2$ | $0.382 \pm 0.017$ | $0.382 \pm 0.017$ | $0.382 \pm 0.017$ | $0.382 \pm 0.017$ |
| $1/Q \times 10^4$ (Mimas) | $1.4 \pm 2.6$ | $1.4 \pm 2.6$ | $1.4 \pm 2.6$ | $1.4 \pm 2.6$ |
| $1/Q \times 10^4$ (Enceladus) | $3.1 \pm 1.2$ | $4.7 \pm 1.2$ | $8.7 \pm 1.3$ | $13.0 \pm 1.3$ |
| $1/Q \times 10^4$ (Tethys) | $1.45 \pm 0.61$ | $1.45 \pm 0.60$ | $1.46 \pm 0.63$ | $1.45 \pm 0.60$ |
| $1/Q \times 10^4$ (Dione) | $3.9 \pm 2.2$ | $3.6 \pm 2.2$ | $3.0 \pm 2.2$ | $2.3 \pm 2.2$ |
| $1/Q \times 10^3$ (Rhea) | $3.67 \pm 0.88$ | $3.67 \pm 0.87$ | $3.68 \pm 0.90$ | $3.68 \pm 0.87$ |
| $1/Q \times 10^2$ (Titan) | $1.8 \pm 1.5$ | $1.8 \pm 1.5$ | $1.8 \pm 1.5$ | $1.8 \pm 1.5$ |

Figure 3 shows a possible orbital evolutionary history for Saturn's moons in the resonance locking framework, using a migration timescale for all moons of $t_{\text{tide}} = 3 t_{\text{Sa}}$, where $t_{\text{Sa}}$ is the changing age of Saturn (see supplementary material for more details). This simple model is roughly consistent with our data, and it incorporates the fact that we expect shorter migration time scales in the past when Saturn was evolving more quickly. Our model implies that Titan's semi-major axis has likely increased by a factor of a few over the age of the solar system, much farther than prior expectations. The substantial migration of Titan may explain how it was able to capture Hyperion into mean motion resonance [24], and a previous resonance crossing with Iapetus may explain the latter's eccentricity and inclination [25]. Due to the changing values of $Q$ associated with resonance locking, Saturn likely had larger $Q$ values in the past, such that moons migrated more slowly than they would by assuming a constant $Q$. Hence, our results reconcile the rapid migration of the inner moons (and large tidal heating rate of Enceladus) with an age of at least a few Gyr, though we still tentatively conclude the inner moons could have formed well after Saturn (supplementary information).

While dissipation of the dynamical tide within Saturn's convective envelope or stably stratified core[26] seems to be the most significant mechanism of tidal friction, dissipation of the equilibrium tide must also contribute. Assessing tidal dissipation in a solid core using a two-layer model, [27] showed that a large range of effective equilibrium tidal dissipation (denoted by $Q_e$) is possible depending on geophysical parameters. Looking at Figure 1 and Table 1, we constrain $Q_e \gtrsim 5000$, but it remains possible that the migration of Mimas, Tethys, and Dione are driven by equilibrium tidal dissipation within Saturn's core or envelope. While our results indicate a small $Q$ for Enceladus that favors an active resonance lock, it is possible that its migration is due to equlibrium tidal dissipation, though this requires an extremely low current dissipation rate (a few GW) within Enceladus. Future astrometric measurements will help reduce the error bar on $Q_e$, clarifying the different energy dissipation mechanisms at work, and allowing for a more accurate picture of the dynamical evolution of Saturn's moons.

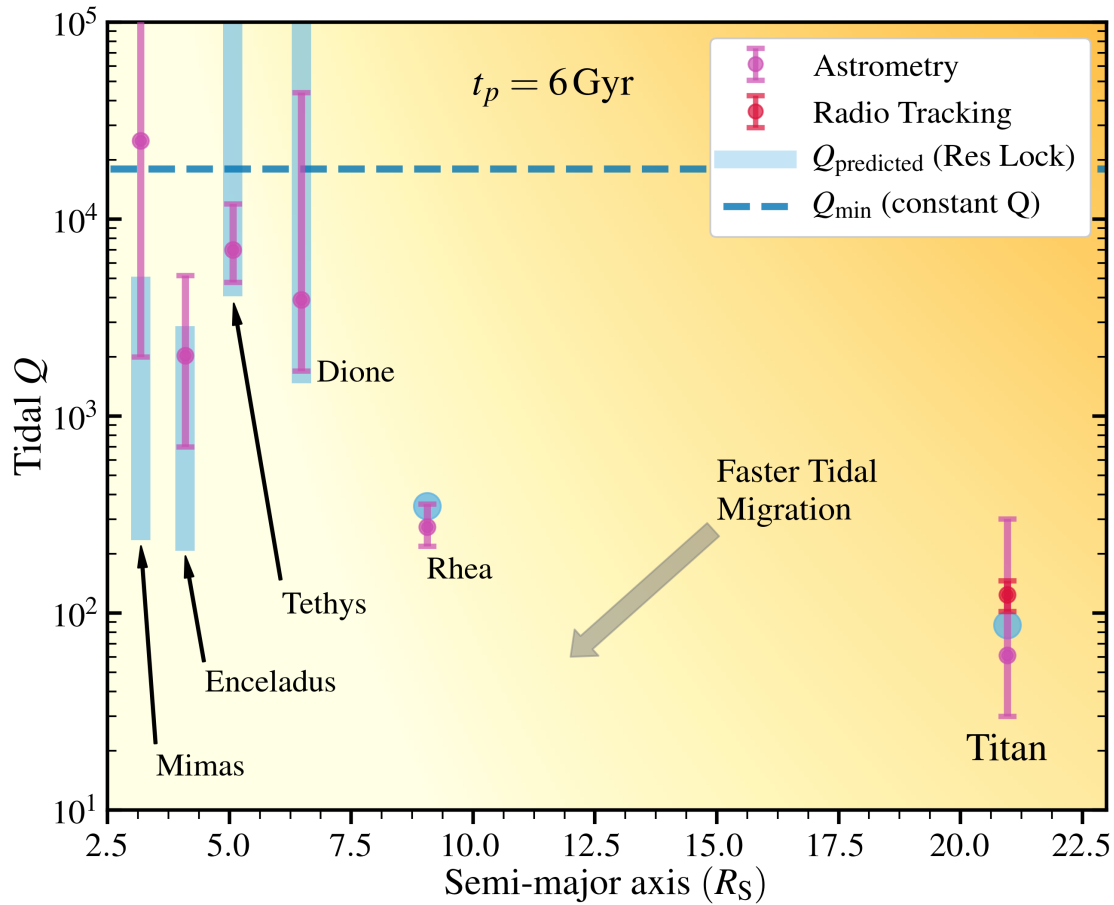

Figure 1: -Saturnian tidal quality factor-

Effective tidal quality factor $Q$ of Saturn for the tidal bulge raised by each of its moons measured in this work. Purple points are measurements from astrometric solutions, with $3\sigma$ error bars, which extend to encompass results from each of the considered energy dissipation rates in Enceladus. Titan's red point is measured using *Cassini* radio tracking data. Blue shaded regions are the predicted tidal quality factors from a resonance locking model with a Saturn evolution time of $t_p = 6$ Gyr. The vertical extent of the blue bars of Mimas, Enceladus, Tethys, and Dione is due to their mean-motion resonances and is determined by their relative migration rates [2]. The horizontal dashed line is the minimum value of $Q$ that allows for coeval formation of Mimas and Saturn, assuming $Q$ is constant [23]. The background is shaded by relative migration timescale, with faster migration in the lower left and slower migration in the upper right.

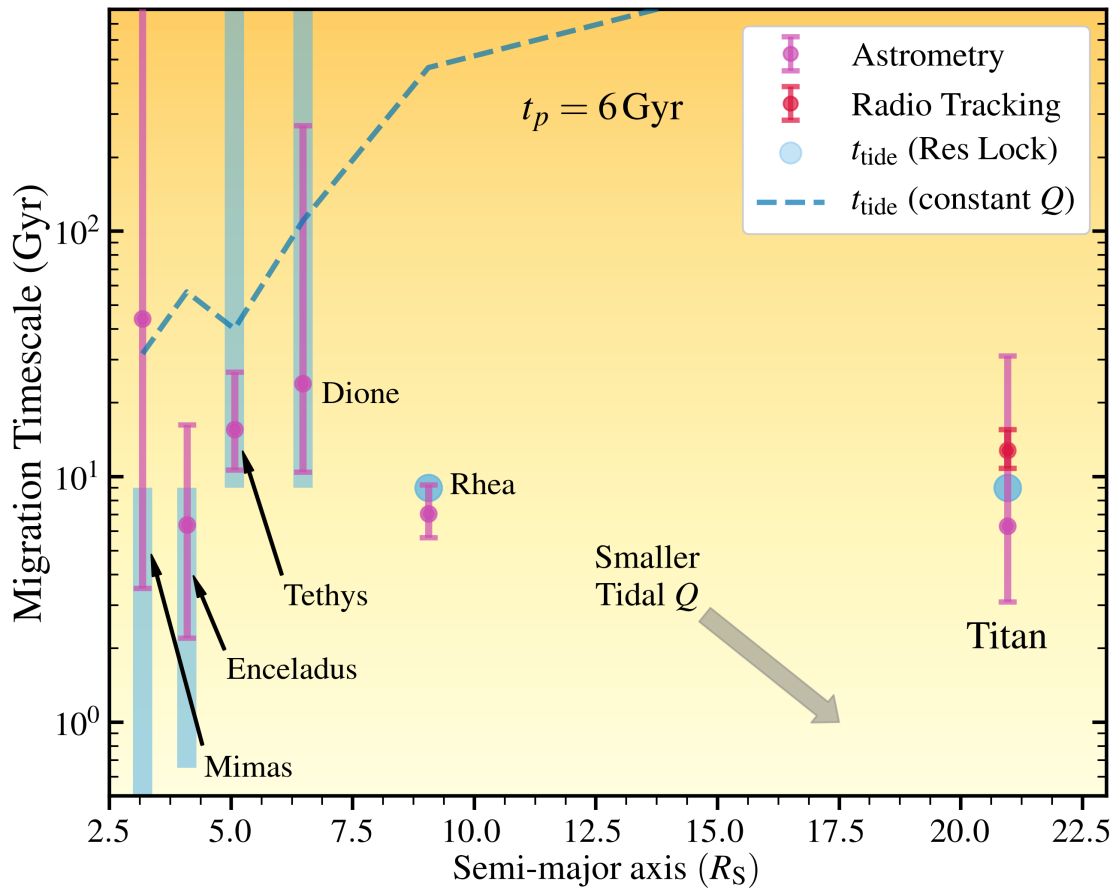

Figure 2: - Tidal migration timescale -

Tidal migration timescale for each of Saturn's moons based on the measurements from Figure 1, in the absence of mean-motion resonances. Blue points show the same resonance locking model as Figure 1, and the background is again shaded by relative migration timescale. The actual migration timescales of Mimas and Enceladus may be longer than the measured values because of mean-motion resonances with Tethys and Dione, whose actual migration timescales may be shorter because they are pushed out by Mimas and Enceladus. The tidal migration timescale with a constant $Q$ (blue dashed line) corresponds to $Q = 1.8 \times 10^4$ as in Figure 1. The migration timescale of each moon is within a factor of $\approx 2$ of 10 Gyr.

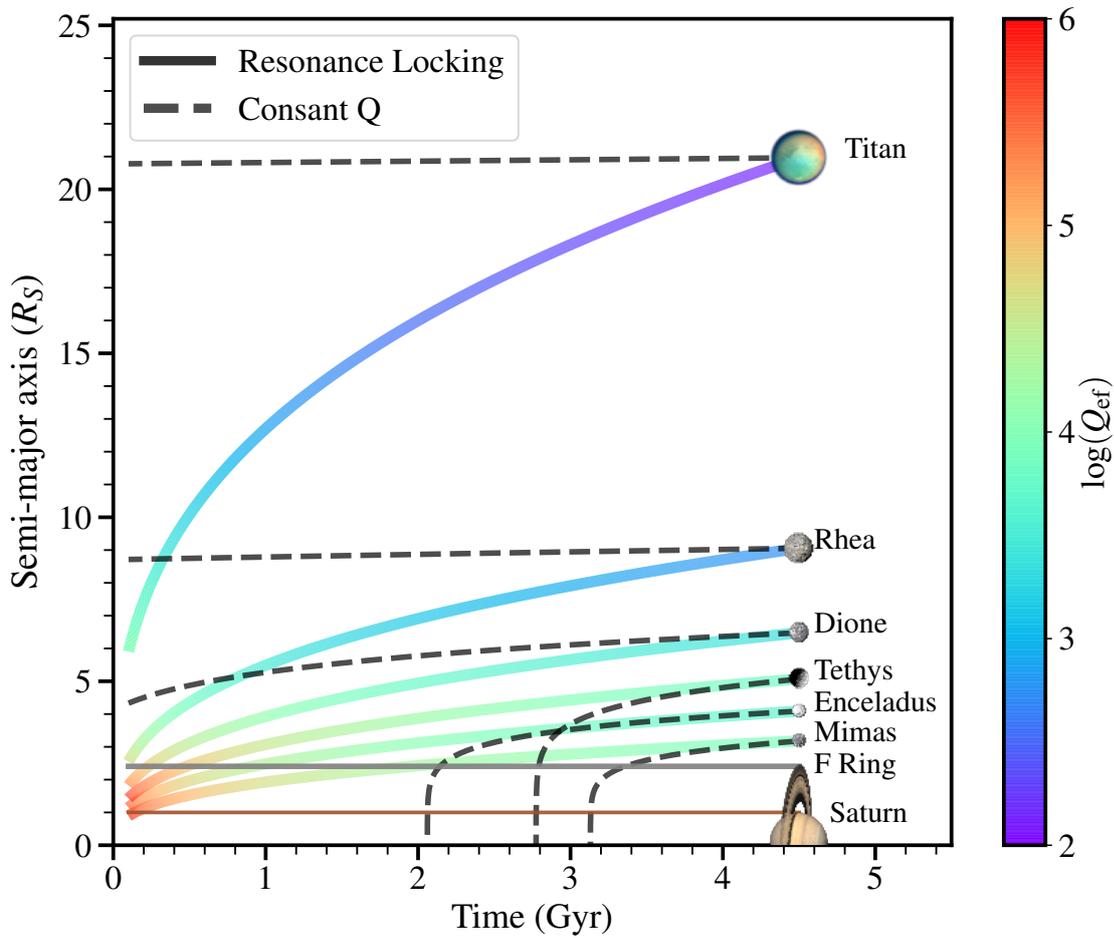

Figure 3: - Moon orbital evolution -

A possible evolutionary history of the orbital distance of Saturn's moons as a function of time, for both a resonance locking model with inertial waves (solid colored lines) and a constant $Q = 5000$ model (black dashed lines). While mean-motion resonances (not accounted for here) can alter these histories, these models illustrate the qualitatively different behaviors of resonance locking and constant $Q$ models. The resonance locking models are shaded by the effective tidal quality factor, $Q_{\text{ef}}$, at a given moment in time. Our results indicate that Titan and Rhea have migrated farther than previously expected, while the inner moons may have a markedly different history than they would in constant $Q$ models.

**Methods**

**\*** Radiometric data selection and calibration

To measure a precise orbit of Titan, we analyzed the *Cassini* radiometric data acquired during 10 close encounters with Titan (T11, T22, T33, T45, T68, T74, T89, T99, T110, and T122) throughout the *Cassini* mission. To increase the sensitivity we selected only the encounters with data coverage around the closest approach.

The main observable used in the reconstruction of *Cassini*'s trajectory is the spacecraft range rate, obtained from the Doppler shift of a microwave carrier transmitted from ground at X-band (7.2 GHz) and sent back coherently at both X- (8.4 GHz) and Ka-band (32.5 GHz). Doppler observables were integrated over a count time of 60 s. In addition, in order to study the orbital evolution of Titan, we used range data at X-band.

We preferred two-way Doppler data over three-way, because of the intrinsic higher stability. When two-way observables were unavailable, we used three-way, adding in our filter the necessary bias to correct for possible DSN inter-station clock offset. When available, X/Ka measurements were preferred over X/X, as they are less sensitive to the dispersive effects, like Earth's ionosphere and solar and interplanetary plasma. We corrected the tracking data for the effects caused by the Earth's troposphere and ionosphere, using Global Positioning System data and microwave radiometer data, when available. Tracking data acquired at ground station elevations lower than 15 degrees were discarded in order to avoid errors due to inaccurate calibration of the tropospheric and ionospheric induced delays. Furthermore, we generated corrections to take into account the additional Doppler shift induced by the spin of the *Cassini* spacecraft.

**Dynamical model**

The dynamical model included the relativistic point-mass gravitational acceleration from the Sun, the planets, the Moon, Pluto, and the main Saturn satellites. In addition, the setup included the gravity field of Saturn and its planetary rings resulting from the analysis of data from the Grand Finale orbits [16]. Saturn's response to the tides raised by Titan was modelled using a complex Love number $k_2$. Furthermore, the model included the following non-gravitational accelerations for *Cassini*: the solar radiation pressure, the drag induced by the upper-layer of Titan's atmosphere, and the acceleration due to the non-isotropic thermal emission, mainly generated by the three onboard Radioisotope Thermoelectric Generators.

No constraints were applied to the estimation of the global parameters, because the a priori uncertainties were chosen to be at least one order of magnitude larger than the obtained formal uncertainties or the formal uncertainty currently available from the literature[4,16]. Besides global parameters, we estimated also local parameters, which means that they affect only a single arc. For each encounter, they include the initial state of *Cassini*, the drag perturbation during the low-altitude flybys, the low gain antenna phase-centre position during T110, constant Doppler bias for the three-way passes, and constant range biases per station and pass. The a priori uncertainties for *Cassini's* position and velocity were 20 km and 0.2 m/s, respectively.

**\*** Measurement of Titan's orbit

The inclusion of the Saturn's tidal dissipation at Titan's frequency to our dynamical model allowed for a fit to the noise level, as shown by the range rate residuals (Figures 4 and 5). The estimated gravity field of Titan and Saturn are both fully compatible with the latest measurements published by the *Cassini* RS team [16,17].

The radiometric data have proven to be very sensitive to the imaginary component of Saturn's Love number, $Im(k_2)$, that drives tidal migration. During each encounter, we can accurately measure the relative position of the *Cassini* spacecraft with respect to Titan. In addition, outside the sphere of influence of Titan, we are sensitive to the relative position of the spacecraft with respect to the gas giant. As a result, the analysis of the data acquired during all close encounters provides information on the relative position of Titan with respect to Saturn. The formal uncertainty in the Titan's measured position is 3 m in the radial direction, almost constant during the timespan of the *Cassini* mission, while the positional uncertainty in the transverse direction has a minimum of 4 m in January 2010, and then grows almost linearly to a maximum of 40 m.

The imaginary part of Saturn's Love number at Titan's frequency causes an orbital migration of the satellite in the radial and transverse position with respect to Saturn. During the timespan of the *Cassini* mission, the expected radial migration given the estimated value of $Im(k_2)$ is about 4 m, which is close to the limit of the data sensitivity. However, Titan's outward migration causes its orbital mean motion to decrease. The secular drift of the mean longitude of Titan increase quadratically with time, reaching a maximum value of 150 m during the observations timespan. This distance is much larger than our formal uncertainty, making the effect of the dissipation clearly observable.

\* Robustness of the solution The stability of our solution has been assessed by carrying out several tests, including changing the values of the less observable $Re(k_2)$, and the use of different ephemerides to take into account possible changes in the orbits of the other Saturn satellites. In all cases, the estimated values were compatible with the reference solution within $1\sigma$, and they offered range-rate residuals of very similar quality.

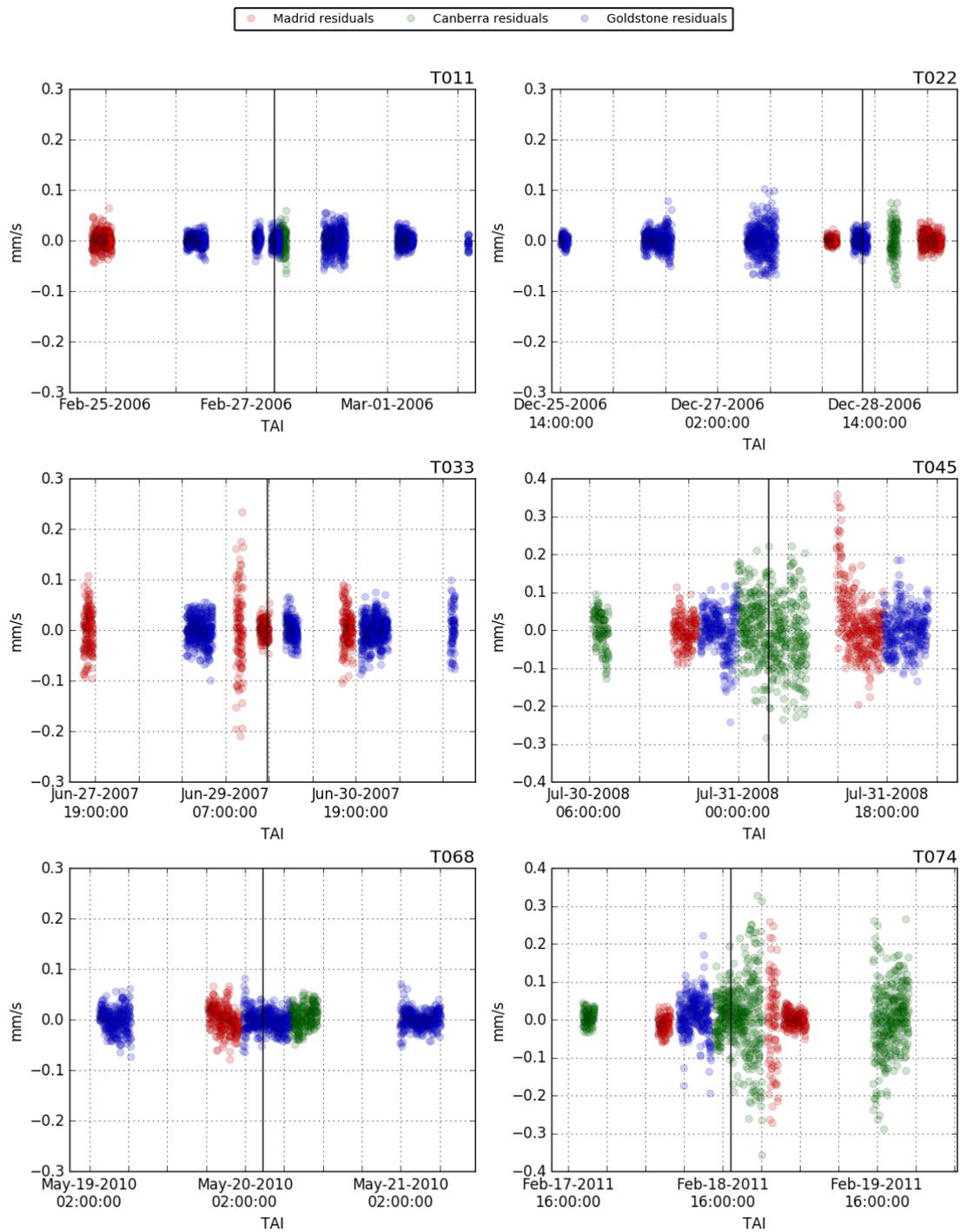

Figure 4: Range-rate residuals integrated over 60s for the T11, T22, T33, T45, T68, and T74 flybys. The vertical line corresponds to the close encounter with Titan.

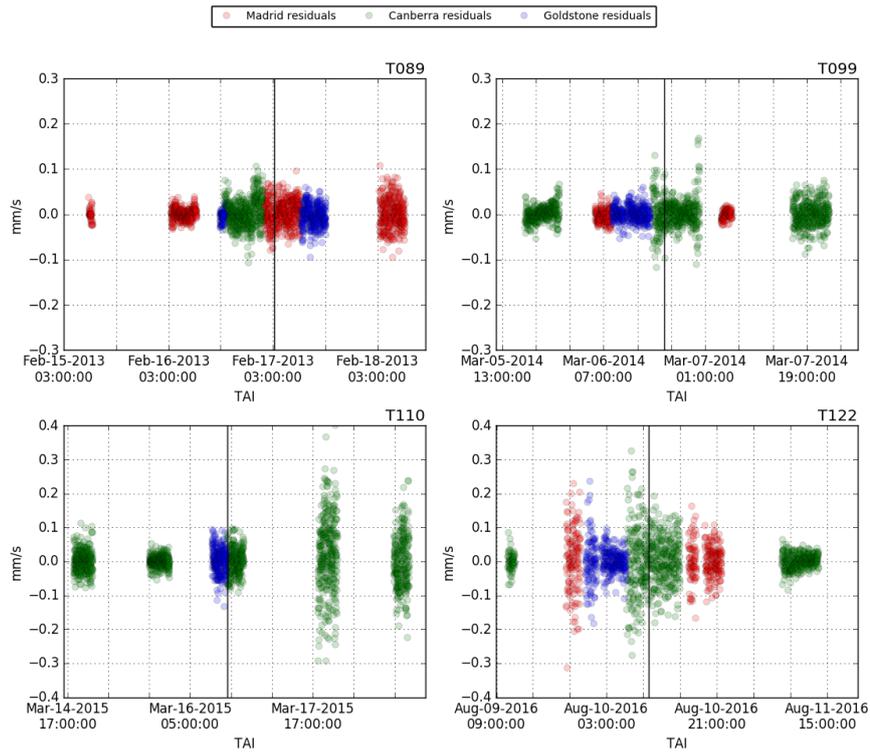

Figure 5: Range-rate residuals integrated over 60s for the T89, T99, T110, and T122 flybys. The vertical line corresponds to the close encounter with Titan.

Another possible source of error may come from a time variation of $J_2$. In particular, a linear variation of $J_2$ produces a quadratic drift on the mean longitude, similar to the effects of Saturn's dissipation. Hence, in order to check the effect of this parameter on our results, we introduced a secular variation of Saturn's $J_2$ equal to $4.42 \times 10^{-13}\,\text{day}^{-1}$, corresponding to the upper limit estimated by the astrometric fit (see Section ). The estimated value of $Im(k_2)$ remained compatible with our nominal solution within $1\sigma$. In the analysis performed using radiometric observables, dissipation within Titan has been neglected. To check the robustness of the obtained solution to this effect, we introduced the $Im(k_{2,Ti})$ of Titan in the list of parameters to solve for, retrieving a solution which is statistically equivalent. In this case, we found the value of $Im(k_{2,Ti})$ of Titan (causing tidal dissipation within Titan) to be $Im(k_{2,Ti}) = -1.248 \times 10^{-2} \pm 4.479 \times 10^{-2}$, compatible with zero within 1-sigma.

* Estimating the secular variation of Saturn's $J_2$

Since a secular variation of Saturn's $J_2$ would provide a secular effect on the moons' semi-major axis, we tried assessing its possible magnitude using astrometry data. For that matter, we considered Pan's orbit. Indeed, Pan orbits extremely close to Saturn and is not involved in any orbital resonance. Moreover, its small size allows for an almost perfect determination of its center of mass from ISS images. Lastly, its small mass does not allow for a significant dynamical coupling with the Saturn's rings. Hence, Pan is a perfect candidate for probing a secular effect on $J_2$. The dynamical modeling we considered here is similar to the one presented in the main text, except that Daphnis was added as a perturber, and the orbits of the coorbitals as well as Methone and Pallene were discarded. All the moons (except Pan itself) were forced from ephemerides. We solved for the initial state vector of Pan, the polar orientation and precession of Saturn, and Saturn's gravity coefficients $J_2$, $J_4$, $J_6$, and $\dot{J}_2$.

To check the potential influence of the choice of moons' ephemerides on the estimation of Saturn's $J_2$ from Pan's orbit, we considered three independent fits. The first one introduced sat382 and sat360xl for the ephemerides of the inner moons and the main moons respectively. The second one introduced sat382 for the inner moons and NOE-6-2018-MAIN ephemerides. These last ephemerides correspond to the solution in the present paper for assuming 33GW of tidal dissipation wihtin Enceladus. Our third test considered NOE-6-2018-inner-eph4 and NOE-6-2018-MAIN ephemerides for the inner and main moons, respectively. We obtained for these three independedt fits the solutions $\dot{J}_2 = (2.3 \pm 4.4) \times 10^{-13}, (2.0 \pm 4.4) \times 10^{-13}, (-1.0 \pm 4.2) \times 10^{-13}$ day$^{-1}$. Hence, no signal could be detected associated with Saturn's $\dot{J}_2$, up to $4.4 \times 10^{-13}$ day$^{-1}$.

**∗** Testing the influence of the inner moons' chaotic motion

We tested the influence of chaos affecting the inner moons' system on our results by developing five different sets of ephemerides. The first three sets were obtained after fitting our model of the inner moons presented in [4] to *Cassini* data, and then extrapolating their orbit over more than a century to cover the full time span considered here. The first set introduced a constant step size in the integration, the second one added extra-precision in the compiling script and the last one introduced a variable step size. The fourth set of ephemerides considered extra precision with variable step size and added *Hubble Space Telescope* data in the fit. The last set was obtained similarly but adding *Voyager* data. While the five sets provide similar ephemeride quality during the *Cassini* era, they start diverging significantly before 2000. Still we found that none of the five ephemerides tested changed our results, as the differences were significantly below the error bar of the measurements. This was somewhat expected since the influence of inner moons is rather small, even on Mimas. Hence, their influence falls way below the accuracy of the measurements (*Cassini* data excluded). Still, it was found that their global influence was important to fit properly the Mimas-Tethys long term libration with *Cassini* data. Moreover, Jacobson (2014) showed that a proper fit of Methone requires consideration of the influence of Pandora on Mimas' orbit.

**Data Availability**

The astrometric datasets generated during and/or analysed during the current study are available from the corresponding author on reasonable request.

The Cassini tracking data and the ancillary information used in this analysis are archived at NASAs Planetary Data System (https://pds.nasa.gov).

**Code availability**

All Astrometric data derived from ISS-images can be reproduced using our CAVIAR software available under Creative Commons Attribution-NonCommercial-ShareAlike 4.0 International License here: www.imcce.fr/recherche/equipes/pegase/caviar

The MONTE space navigation code was obtained through a license agreement between NASA and the Italian Space Agency; the terms do not permit redistribution. MONTE licenses may be requested at https://montepy.jpl.nasa.gov/.

The availability of NOE software is limited due to NASA restrictions.

**Acknowledgments**

V.L.'s research was carried out at the Jet Propulsion Laboratory, California Institute of Technology, under a contract with the National Aeronautics and Space Administration. This work has been supported by the ENCELADE team of the International Space Science Institute (ISSI). Support for this work was provided by the Italian Space Agency (LGC, MZ, PT, and DM) through the Agreement 2017-10-H.O in the context of the NASA/ESA/ASI Cassini/Huygens mission. JF's research is funded in part by a Rose Hills Innovator Grant. N.C. and C.M. were supported by the UK Science and Technology Facilities Council (Grant No. ST/M001202/1) and are grateful to them for financial assistance. N.C. thanks the Scientific Council of the Paris Observatory for funding. Q.Z.'s research was supported by the National Natural Science Foundation of China (Grant No. 11873026).

# 1 Supplementary Information

**Resonance Locking Dynamics** Our observations indicate that resonance locking operates between Saturn and Titan, so it is worth investigating different types of resonance locking that could be operating.

### 1.0.1 Accounting for Titan's Angular Momentum

Prior work [2] neglected the angular momentum (AM) of moons because they are usually very small compared to that of the planet. Titan's large mass and semi-major axis, however, means that its orbital moment of inertia is roughly 1/3 that of Saturn, and is non-negligible. To include this effect, we examine the system's total AM $J$,

$$J = J_{\text{Sa}} + J_{\text{Ti}}$$

$$\simeq I_{\text{Sa}}\Omega + M_{\text{Ti}}\sqrt{GM_{\text{Sa}}a_{\text{Ti}}}, \qquad (2)$$

where $\Omega$ is Saturn's angular rotation frequency, and $I_{\text{Sa}}$ is its moment of inertia. We neglect the AM of other moons, which are negligible in comparison. Taking the time derivative of equation 2, assuming the total AM and masses are conserved, and neglecting effects of other moons, we have

$$0 = I_{\text{Sa}}\dot{\Omega} + \dot{I}_{\text{Sa}}\Omega + \frac{1}{2}M_{\text{Ti}}\sqrt{GM_{\text{Sa}}a_{\text{Ti}}}\frac{\dot{a}}{a}. \qquad (3)$$

As Saturn evolves, its moment of inertia and rotation rate will change. Defining evolutionary timescales $t_{\text{p}}^{-1} = \dot{\Omega}/\Omega$, and $t_I^{-1} = \dot{I}/I$, equation 3 can be rearranged to find

$$t_{\text{p}}^{-1} = -t_I^{-1} - \frac{J_{\text{Ti}}}{2J_{\text{Sa}}}t_{\text{tide}}^{-1}. \qquad (4)$$

The coupled planetary and orbital evolution during a resonance lock is derived in [2]. Inserting equation 4 into equation 12 of [2], we find the resonance locking migration rate of Titan,

$$t_{\text{tide}}^{-1} = \frac{2}{3}\left[\frac{\Omega}{\Omega_{\text{Ti}}}\left(t_\alpha^{-1} + t_I^{-1} + \frac{J_{\text{Ti}}}{2J_{\text{Sa}}}t_{\text{tide}}^{-1}\right) - t_\alpha^{-1}\right]. \qquad (5)$$

Then we have

$$t_{\text{tide}}^{-1} = \frac{2}{3}\left(1 - \frac{I_{\text{Ti}}}{3I_{\text{Sa}}}\right)^{-1}\left[\frac{\Omega}{\Omega_{\text{Ti}}}\left(t_\alpha^{-1} + t_I^{-1}\right) - t_\alpha^{-1}\right]. \qquad (6)$$

The value of $I_\text{Ti}/I_\text{Sa} \sim 1/3$, so we find that including Titan's moment of inertia only produces a $\sim 10\%$ correction to the resonance locking migration rate. Our theoretical predictions neglect this effect because it is smaller than our measurement uncertainties, and smaller than the theoretical uncertainties in parameters such as $t_\alpha$ and $t_I$.

### 1.0.2 Accounting for Coriolis forces

If resonance locking occurs via a tidally excited gravity mode (g mode), Coriolis forces will affect its frequency evolution. Using the traditional approximation, the angular frequency $\omega$ and radial wavenumber $k_r$ of a g mode are related by

$$\omega = \frac{\lambda^{1/2} N}{r k_r}, \tag{7}$$

where $N$ is the Brunt-Väisälä frequency, $\lambda$ is the angular eigenvalue of the g mode, and $\lambda = \ell(\ell+1)$ in the non-rotating limit. Note that in the rotating frame of Saturn, Titan's tidal forcing frequency is $\omega_\text{force} = m(\Omega - \Omega_\text{Ti})$, whree $m$ is the azimuthal mode number. Since $\Omega_\text{Ti} \ll \Omega$, we have $\omega_\text{force} \simeq m\Omega$. For $m=2$ modes, the Coriolis parameter is $\nu = 2\Omega/\omega_\text{force} \simeq 2/m$. We expect $m=2$ modes are the most likely to contribute to tidal interaction, so we expect $\nu \sim 1$.

For a mode of given radial order, the value of $rk_r$ in equation 7 is nearly constant. Taking the time derivative of equation 7, we then have

$$t_\alpha^{-1} \simeq \frac{\dot\lambda}{2\lambda} + \frac{\dot N}{N}. \tag{8}$$

Then using

$$\begin{aligned}
\dot\lambda &= \frac{d\lambda}{d\nu}\frac{d\nu}{dt} \\
&= \frac{d\lambda}{d\nu}\left(\frac{2\dot\Omega}{\omega} - \frac{2\Omega\dot\omega}{\omega^2}\right) \\
&= \lambda\frac{d\ln\lambda}{d\ln\nu}\left(\frac{\dot\Omega}{\Omega} - \frac{\dot\Omega - \dot\Omega_{\rm Ti}}{\Omega - \Omega_{\rm Ti}}\right) \\
&= \lambda\frac{d\ln\lambda}{d\ln\nu}\left(\frac{-3\Omega_{\rm Ti}\Omega t_{\rm tide}^{-1} - 2\Omega_{\rm Ti}\Omega t_{\rm p}^{-1}}{\Omega(\Omega - \Omega_{\rm Ti})}\right).
\end{aligned} \qquad (9)$$

Inserting equation 9 into equation 8, and inserting that into the resonance locking condition of equation 12 of [2], we find

$$t_{\rm tide}^{-1} = \frac{2}{3(1 + d\ln\lambda/2d\ln\nu)}\left[\frac{\Omega - \Omega_{\rm Ti}}{\Omega_{\rm Ti}}\frac{\dot N}{N} - \left(\frac{\Omega}{\Omega_{\rm Ti}} + \frac{1}{2}\frac{d\ln\lambda}{d\ln\nu}\right)t_{\rm p}^{-1}\right]. \qquad (10)$$

Near $\nu \sim 1$, we expect $d\ln\lambda/d\ln\nu \sim 1$, so inclusion of Coriolis forces has a modest effect on the resonance locking dynamics of gravity modes.

Since $\Omega_{\rm Ti} \ll \Omega$, we find the resonance locking migration rate of equation 10 due to gravity modes approximately scales as

$$t_{\rm tide}^{-1} \propto \frac{\Omega}{\Omega_{\rm moon}}. \qquad (11)$$

Resonance locking with g modes thus predicts *faster* migration for moons with larger semi-major axes. The observational results do not appear to show this trend, and instead it appears that $t_{\rm tide}$ is approximately constant for each of Saturn's moons. This may favor resonance locking with inertial waves, discussed below.

### 1.0.3 Resonance locking with inertial waves

Following [2], "resonances" with inertial waves occur near frequencies where inertial waves are focused onto attractors [8] and the energy dissipation rate is much larger. In the inertial frame, the resonance lock condition is

$$-m\Omega_{\text{moon}} = (c - m)\Omega,  \tag{12}$$

where $c < 2$ is a constant determined by the internal structure of Saturn. Note that if $c$ remains constant as Saturn's internal structure evolves, then migration in resonance lock with inertial waves requires

$$\frac{\dot{\Omega}_{\text{moon}}}{\Omega_{\text{moon}}} = \frac{\dot{\Omega}}{\Omega}. \tag{13}$$

This translates to

$$t_{\text{tide}}^{-1} = -\frac{2}{3} t_p^{-1}. \tag{14}$$

In this very simple model, resonantly locked moons all have *the same* migration timescale outward, regardless of their semi-major axis or mass. However, in order for resonance locking to occur, the planet's spin frequency must *decrease* with time. This is contrary to what is expected if the planet is slowly gravitationally contracting, which typically results in spin up with time.

In the more general case when $c$ evolves with time, we have

$$\frac{\dot{\Omega}_{\text{Ti}}}{\Omega_{\text{Ti}}} = -\frac{\dot{c}}{m - c} + \frac{\dot{\Omega}}{\Omega}, \tag{15}$$

leading to

$$t_{\text{tide}}^{-1} = \frac{2}{3} t_c^{-1} - \frac{2}{3} t_p^{-1}, \tag{16}$$

where $t_c = \dot{c}/(m - c)$.

As pointed out by [2], in the limit of constant spin, resonance locking with inertial modes requires that $t_c$ be positive. Since the quantity $(m - c)$ is positive for $m = 2$, this requires that $\dot{c}$ is positive, i.e., the frequency of the inertial wave attractor (as measured in the rotating frame) must be increasing. While the value of $\dot{c}/(m - c)$ depends on the internal structural evolution of Saturn, we might expect this quantity to be similar for each moon caught in a resonance lock with inertial waves. We would thus expect a similar value of $t_{\rm tide}$ for each moon, though perhaps with order unity variance. This is consistent with the observational results of the main text. Hence, we suggest that resonance locking with inertial waves (rather than g modes) is the most likely explanation for the observed moon migration rates. In this case, we do not expect to be able to detect perturbations to Saturn's gravity field due to resonantly excited gravity modes or inertial waves [28].

**Orbital Evolution of Moons** Our results in Figure 2 demonstrate that Saturn's moons currently have a similar migration time scale of order $t_{\rm tide} \sim 10 \, {\rm Gyr}$. Here we explore models of the long-term orbital evolution of the moons. In the resonance locking theory, the migration time scale is determined by the time scale on which the planet's structure and spin rate evolve. More detailed modeling of the coupled evolution of Saturn's interior structure, and the consequential resonance locking migration timescale, should be performed to develop reliable predictions. However, we note that planets generally evolve much faster (e.g., by cooling and contracting, [29–32]) when they are young. At any given time since formation, we expect the evolution timescale of the planet (and hence the value of $t_{\rm tide}$) to be comparable to its age. Hence, we expect orbital evolution of the form

$$\frac{1}{t_{\rm tide}} = \frac{1}{a}\frac{da}{dt} \approx \frac{B}{t}, \qquad (17)$$

where $t$ is Saturn's age and $B$ is a constant of order unity that is determined by the precise rate at which Saturn's inertial wave frequencies are evolving. While $B$ is difficult to predict from first principles, Figure 2 shows that the moons all have a current migration timescale of roughly $t_{\rm tide} \sim 13\,{\rm Gyr}$, corresponding to $B \sim 1/3$, though we note $B$ may vary slightly for different moons, and at different times.

Nonetheless, we can attempt to solve for the approximate orbital evolution of the moons. Integrating equation 17, we can solve for the orbital distance as a function of time,

$$a = a_0 \left(\frac{t}{t_0}\right)^B, \qquad (18)$$

where $a_0$ and $t_0$ are a moon's current semi-major axis, and Saturn's current age. So, we expect the orbits to expand as a power law function of time. This is very different from the evolution expected from a constant $Q$ as usually assumed. Combining equation 17 with equation 1, we can solve for an effective tidal quality factor $Q_{\rm ef}$ as a function of orbital distance,

$$Q_{\rm ef} = \frac{3k_2}{B} \frac{G^{1/2} M_{\rm moon} R^5}{M_{\rm Sa}^{1/2}} \frac{t_0}{a_0^{1/B}} a^{-13/2 + 1/B}. \qquad (19)$$

So, the effective quality factor is a sensitive function of the semi-major axis of a moon, and should not be assumed to be constant in time or space.

Figure 3 shows an example of the expected orbital evolution of Saturn's moons, using $B = 1/3$ for each moon, which predicts that each moon's orbital distance evolves as $a \propto t^{1/3}$. We do not take mean-motion resonances into account, but we note that for a constant value of $B$, the orbital period ratio of moons remains constant. We solve for the moon's positions from the present day back to an age of 100 Myr, before which resonance theory may start to break down. Figure 3 also shows an orbital evolution for a constant $Q = 5000$, roughly consistent with the inner moons' current orbital expansion rates. We can see that assuming a constant $Q$ produces drastically different orbital evolution, leading to faster expansion of the inner moons' orbits and slower expansion of the outer moons' orbits. Since our results favor resonance locking over constant $Q$ theories, orbital evolution calculations assuming that $Q$ is constant (either in space or in time, e.g., [33]) should be revisited.

We caution that the evolution shown in Figure 3 is a fairly crude model for the long-term evolution of Saturn's moon system. Nonetheless, we expect the resonance locking tracks in Figure 3 to be better approximations than constant $Q$ models. These results indicate that the semi-major axes of Rhea and Titan may have evolved by a factor of a few over the age of the solar system, orders of magnitude more than prior expectations. Figure 3 also indicates that the inner moons (Mimas, Enceladus, Tethys, and Dione) may have formed out of Saturn's rings (i.e., where the tracks cross the gray horizontal line in Figure 3) well after the formation of Saturn, as postulated by several recent works [5,6,34]. However, this result is tentative because it depends on the exact value of $B$ for different moons and whether it evolves with time. More detailed calculations should be performed to answer this question, examining the coupled evolution of Saturn's interior, dynamical tidal response, and moon orbital positions.

**Extra References**  R.A. Jacobson, American Astronomical Society, DDA meeting 45, id.304.05. V. Lainey, B. Noyelles, N. Cooper, N. Rambaux, C. Murray, R.S. Park, *Icarus* (2019), 10.1016/j.icarus.2019.01.026.